\documentclass[unsortedaddress,aps,preprint]{revtex4}
\usepackage{graphicx}
\usepackage{dcolumn}
\usepackage{amsmath}
\usepackage{amssymb}
\usepackage{amsfonts}
\usepackage{bm}
\usepackage{color}
\usepackage[normalem]{ulem}
\usepackage{hyperref}
\hypersetup{colorlinks=true,linkcolor=blue,urlcolor=magenta}
\newcommand{\be}{\begin{equation}}
\newcommand{\ee}{\end{equation}}
\newcommand{\ba}{\begin{eqnarray}}
\newcommand{\ea}{\end{eqnarray}}

\begin{document}
\title{Ultracold bosons on a regular spherical mesh}
\author{Santi Prestipino$^1$\footnote{Email: {\tt sprestipino@unime.it}}}
\affiliation{$^1$Universit\`a degli Studi di Messina,\\Dipartimento di Scienze Matematiche ed Informatiche, Scienze Fisiche e Scienze della Terra,\\Viale F. Stagno d'Alcontres 31, 98166 Messina, Italy}

\begin{abstract}I study the zero-temperature phase behavior of bosonic particles living on the nodes of a regular spherical mesh (``Platonic mesh'') and interacting through an extended Bose-Hubbard Hamiltonian. Only the hard-core version of the model is considered here, for two instances of Platonic mesh. Using the mean-field decoupling approximation, I show that the system may exist in various ground states, which can be regarded as analogs of gas, solid, supersolid, and superfluid. For one mesh, by comparing the theoretical results with the outcome of numerical diagonalization, I manage to uncover the signatures of diagonal and off-diagonal spatial orders in a finite quantum system.
\end{abstract}
\maketitle
\section{Introduction}

Gases of ultracold bosonic atoms loaded in an optical lattice provide the unique opportunity to study quantum many-body effects under controlled conditions~\cite{Bloch,Amico}. To a very good approximation, the atoms can be described by a Bose-Hubbard (BH) Hamiltonian~\cite{Fisher} whose parameters can be tuned by laser light~\cite{Jaksch,Greiner}. By changing the configuration of the lasers, many lattice geometries can be explored, making optical lattices a powerful and versatile tool.

In a system at zero temperature ($T=0$), thermal fluctuations are frozen out and quantum fluctuations prevail. These microscopic fluctuations can induce phase transitions in the ground state of a many-body system, driven by a non-thermal control parameter, such as chemical potential, magnetic field, or chemical composition. As a concrete example, consider a dilute gas of bosons at temperatures low enough that a Bose-Einstein condensate is formed. The condensate is described by a wave function consisting of every particle in the same state spread over the entire volume of the system, and typically exhibits superfluidity. An interesting situation appears when the condensate is subject to a lattice potential in which particles can move from one lattice site to the next only by tunneling. If the lattice potential is increased smoothly, the system remains in the condensed phase as long as the repulsion between atoms is small compared to the tunnel coupling (assuming, by the way, that the range of the repulsion is much smaller than the lattice spacing). In this regime, where the tunneling term dominates the Hamiltonian, a delocalized wave function still minimizes the total energy of the many-body system. But when the strength of the repulsion becomes large compared to the tunnel coupling, the total energy is made minimum when each lattice site is occupied by the same number of atoms; as a result, phase coherence is lost and the system becomes insulating. The addition of a longer-range repulsion will make the phase behavior richer, with the possibility of a non-superfluid density wave and a supersolid ground state where crystalline order coexists with superfluid behavior (see, e.g., Ref.~\cite{vanOtterlo}). In experiment, a way to prepare ultracold gases of long-range interacting bosons is to use atoms (such as chromium~\cite{Griesmaier} or dysprosium~\cite{Lu}) and molecules having a large magnetic or electric dipole moment.

The usual BH model predicts a $T=0$ phase transition from a superfluid phase to a Mott insulator phase as the ratio of the hopping matrix element between adjacent sites ($t$, in absolute terms) to the on-site interaction ($U$) is reduced. The overall number density of particles is controlled via a chemical-potential parameter $\mu$. As $\mu$ grows at fixed $t$ and $U$, the lattice becomes increasingly filled with particles, but this can only occur outside the Mott regions since the insulator phase is incompressible~\cite{Fisher}. The BH model has been studied in many lattice geometries and with several techniques (mean-field theory~\cite{Rokhsar,Krauth,Sheshadri}, perturbation theory~\cite{Freericks,vanOosten,Schroll,dosSantos,Kuebler}, and quantum Monte Carlo simulation~\cite{Batrouni,Capogrosso-Sansone}, to name but the most commonly employed). When a further repulsion $V$ is introduced between nearest-neighbor atoms (``extended BH model''), new phases may arise, {\em in primis} a supersolid phase~\cite{Goral,Kovrizhin,Sengupta,Yamamoto,Pollet,Ng,Iskin,Kimura,Ohgoe}. Another variant of the BH model is hard-core bosons, where site occupancy is restricted to zero or one, corresponding to taking the $U\rightarrow\infty$ limit~\cite{vanOtterlo,Wessel,Kurdestany,Zhang,Yamamoto2,Gheeraert}.

I hereafter present the results of yet another investigation of the extended BH model, now choosing a finite graph as hosting space for bosons. Even though clearcut phase transitions (i.e., thermodynamic singularities) cannot occur in a few-particle system, a convenient choice of boundary conditions may alleviate the difference with an infinite system, making the study of a finite quantum system valuable anyway. A practical solution is to use spherical boundary conditions (SBC), which have often been exploited in the past to discourage long-range triangular ordering at high density~\cite{Post,Prestipino,Prestipino2,Prestipino3,Vest,Guerra}. On the other hand, SBC make it possible to observe novel forms of ordering, viz. into regular polyhedral structures, that are simply unknown to Euclidean space --- see Refs.~\cite{Franzini,Prestipino4,Prestipino5}. An added value of a spherical mesh of points is the possibility to vary the site coordination while keeping the overall geometry strictly two-dimensional. Bosons confined in spherical (bubble) traps have been produced experimentally~\cite{Zobay,Garraway} and are going to be studied soon under microgravity conditions~\cite{Elliott,Lundblad}. In the present study, the extended BH model is considered on a finite mesh of points homogeneously distributed on the unit sphere, i.e., coincident with the vertices of a regular polyhedron inscribed in the sphere (we may call it a {\em Platonic mesh}). Despite consisting of a finite number of nodes, a regular spherical mesh shares an important feature in common with an ordinary lattice, namely all sites are equivalent; for this reason, the phase behavior of particles living on a Platonic mesh would not deviate much from that of an infinite system (I will check this intuition in a particular case).

The plan of the paper is the following. After introducing the models in Section 2, the choice of the underlying  mesh is discussed in detail, giving priority to those regular grids where a subset of nodes form a mesh that is also regular. Then, in Section 3 a mean-field (MF) analysis of the ground state is carried out for all the models considered; in one case, the indications of theory are validated against the results of exact diagonalization (Section 4). From this comparison we find what are the artifacts of the MF approximation as applied to a finite system. Finally, some concluding remarks are given in Section 5.

\section{Particle models on a spherical mesh}

\subsection{Classical models}

To illustrate the main idea, let it initially be considered a system of classical point particles defined on the sites of a cubic mesh stretched over the surface of the unit sphere. Each of the eight nodes of the mesh has three nearest neighbors (NN, at chord distance $r=2/\sqrt{3}$) and three next-nearest neighbors (NNN, at chord distance $2\sqrt{2/3}$). For simplicity, assume that the occupancy $n_i$ of site $i$ can only be 0 or 1 ($i=1,\ldots,8$). Finally, choose the system Hamiltonian to be $H[n]=V\sum_{\left<i,j\right>}n_in_j$ with $V>0$ (each NN pair in the sum is counted only once). The grand potential $\Omega(T,\mu)$ of this system is
\be
\Omega=-\beta^{-1}\ln\Xi\,\,\,\,\,\,{\rm with}\,\,\,\,\,\,\Xi=\sum_{\{n\}}e^{-\beta(H[n]-\mu\sum_in_i)}\,,
\label{eq-1}
\ee
where $\beta$ is the inverse temperature. At $T=0$ the formula is simpler:
\be
\Omega=\min_{\{n\}}\left(H[n]-\mu\sum_in_i\right)\,.
\label{eq-2}
\ee
Any of the points of absolute minimum in (\ref{eq-2}) represents the actual equilibrium state of the system for that $\mu$. For particles living in a continuous space, minimization of $H-\mu N$ is carried out among a selection of crystalline states that are thought to be relevant based on symmetry considerations (see, e.g., Ref.~\cite{Prestipino6}). In the present case, where the total number of microstates is small ($2^8=256$), the $T=0$ grand potential can be determined exactly for each $\mu$ by a scrutiny of all possible energies. While the mesh is empty for $\mu<0$ and completely filled with particles for $\mu>3V$, in the interval from 0 to $3V$ only half of the nodes are occupied, and these fall at the vertices of a regular tetrahedron. This twofold-degenerate state with checkerboard order can be viewed as the finite-size counterpart of a crystalline phase.

The rationale behind the choice of a cubic mesh is now clear: by introducing a repulsion between occupied NN sites, we promote the occurrence of a Platonic ``crystal'', i.e., the regular tetrahedron (``CT model'', see Fig.\,1 left). There is only one other possibility to obtain a non-frustrated crystal-like state at $T=0$, which is using a regular dodecahedral mesh. We shall see that i) by discouraging the occupancy of NN sites, a cubic ``crystal'' is stabilized for sufficiently small $\mu>0$ (``DC model'', Fig.\,1 center); ii) if the repulsion is extended to embrace NNN sites too, then a tetrahedral ``crystal'' is stabilized in a range of positive $\mu$ values (``DT model'', Fig.\,1 right).

\begin{figure}
\centering
\includegraphics[width=5 cm]{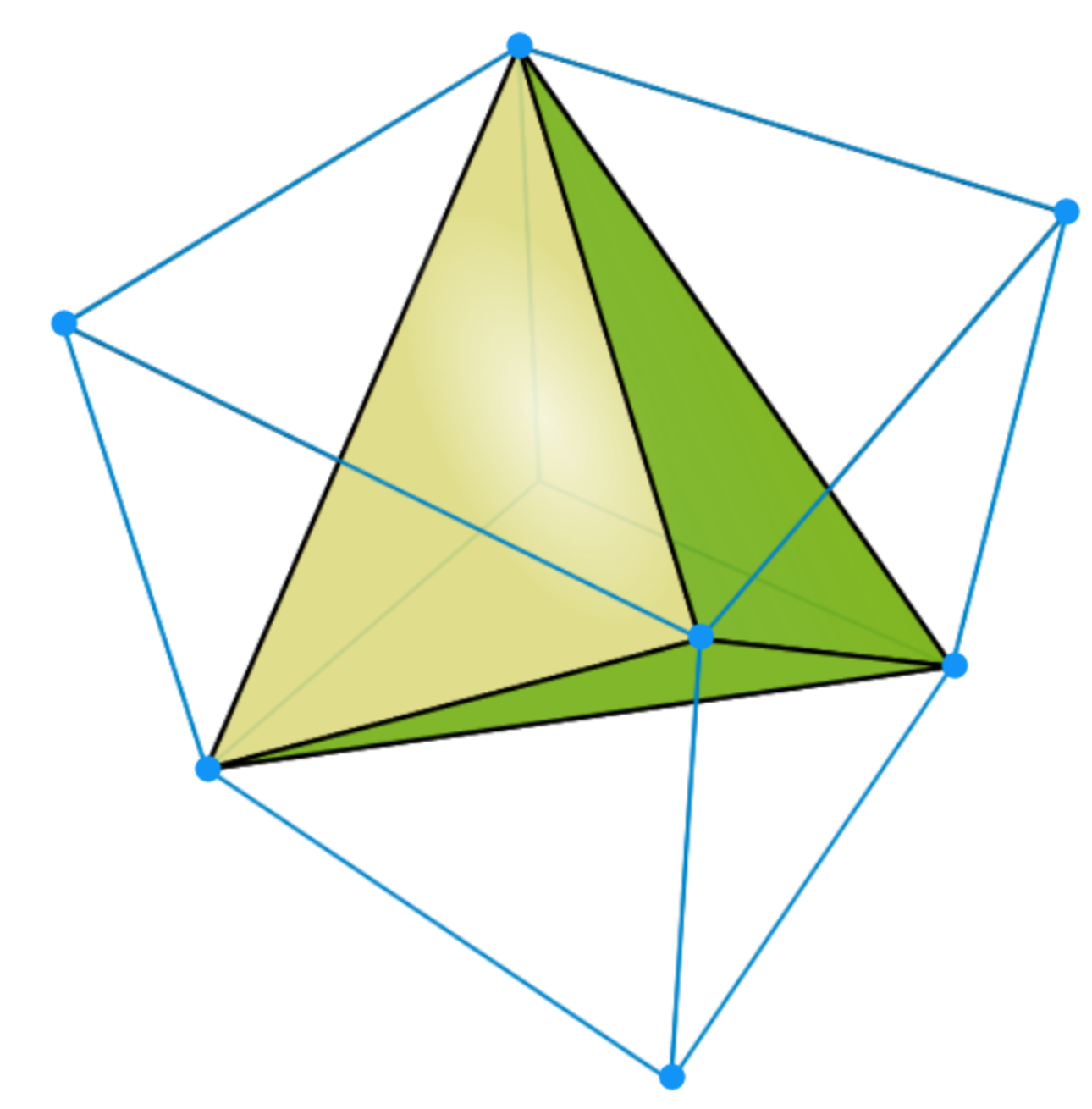}
\includegraphics[width=5 cm]{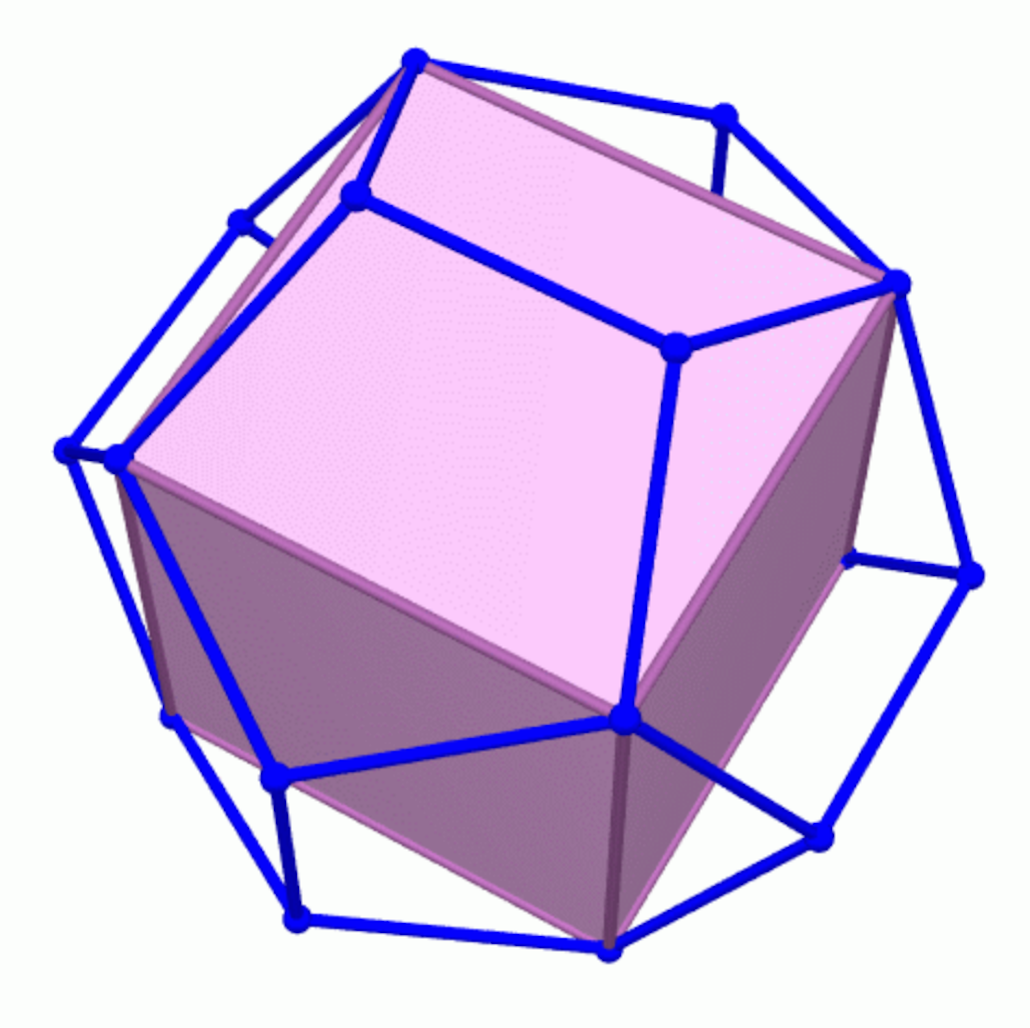}
\includegraphics[width=5 cm]{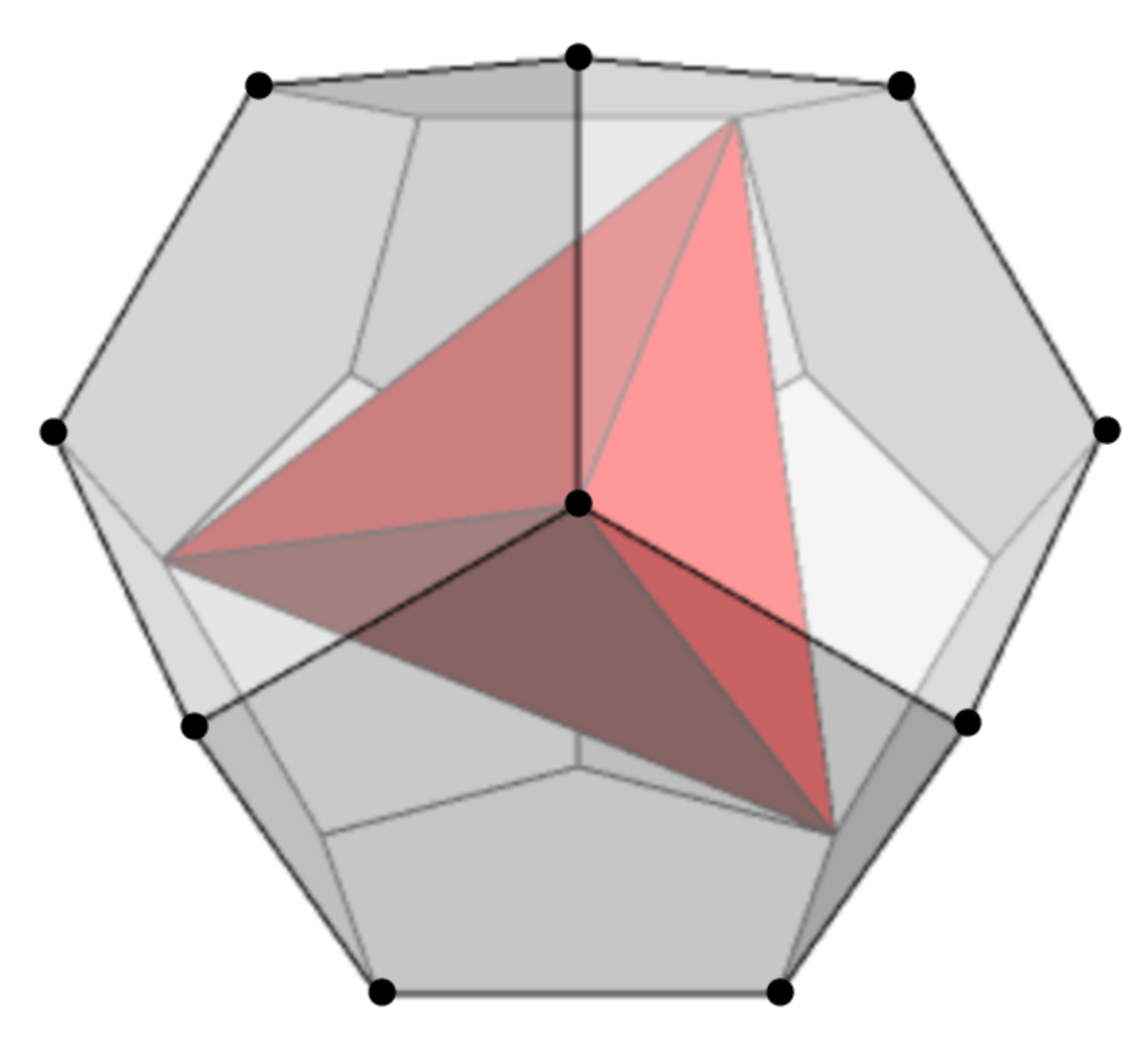}
\caption{The two regular meshes considered in this work (the circumscribed sphere is not shown). The dots (some are hidden) are the sites/nodes of the mesh; their number is 8 for the cubic mesh and 20 for the regular dodecahedral mesh. Left: cubic mesh with a regular tetrahedron inside (CT model). Center: regular dodecahedral mesh with a cube inside (DC model). Right: regular dodecahedral mesh with a regular tetrahedron inside (DT model).}
\end{figure}

For a system of classical hard-core particles defined on a regular dodecahedral mesh, the total number of microstates is $2^{20}=1\,048\,576$. By direct inspection, we see that the minimum-$\Omega$ ``phase'' for the DC model is the empty mesh for $\mu<0$, a cube for $0<\mu<3V/2$ (a fivefold-degenerate state, since there are 5 ways to form a cube with the vertices of a regular dodecahedron), and the complement of a cube (``co-cube'' in the following) for $3V/2<\mu<3V$; finally, for $\mu>3V$ the mesh is completely filled. For the DT model, the mesh is empty for $\mu<0$, filled with particles located at the vertices of a regular tetrahedron for $0<\mu<3V/2$ (a tenfold degenerate state), and completely filled for $\mu>9V$. For $\mu$ between $3V/2$ and $9V$, a different ground state exists for each even number of occupied sites in the range 6 to 16, none of which corresponds to a simple geometric arrangement.

I have plotted in Fig.\,2 the evolution with temperature of a few properties of the CT and DC models. Besides the grand potential $\Omega$, the figure shows the total number of occupied sites ($N$), the total energy ($E$), and the order parameters for tetrahedral ($S_{\rm t}$) and cubic order ($S_{\rm c}$). The latter quantities are defined so as to discriminate the Platonic ``phase'' from the other $T=0$ states. A proper order parameter should be insensitive to the orientation of the polyhedron, hence it can only depend on the relative angles between the vertices vector radii departing from the sphere center: for a regular tetrahedron all these angles are equal to $\alpha=\arccos\{-1/3\}$ (with $\sin\alpha=2\sqrt{2}/3$), while they are $\pi-\alpha,\alpha$, and $\pi$ for a cube. With the idea to penalize configurations that do not match the wanted structure, my choice of the OPs is the following:
\be
S_{\rm t}=\left\{
\begin{array}{ll}
1-k_{\rm t}\sum_{i<j}\left(\cos\theta_{ij}+\frac{1}{3}\right)^2 & \,,\,\,\,N=4\\
0 & \,,\,\,\,N\ne 4
\end{array}
\right.
\label{eq-3}
\ee
and
\be
S_{\rm c}=\left\{
\begin{array}{ll}
1-k_{\rm c}\sum_{i<j}\sin\theta_{ij}\left(\sin\theta_{ij}-\frac{2}{3}\sqrt{2}\right)^2 & \,,\,\,\,N=8\\
0 & \,,\,\,\,N\ne 8\,,
\end{array}
\right.
\label{eq-4}
\ee
where the constants $k_{\rm t}$ and $k_{\rm c}$ are chosen, following the advice in Ref.\,\cite{Errington}, so that $S$ vanishes for a random distribution of angles:
\be
k_{\rm t}=\frac{3}{8}\,\,\,\,\,\,{\rm and}\,\,\,\,\,\,k_{\rm c}=\frac{36}{7}\left(59\pi-128\sqrt{2}\right)^{-1}\,.
\label{eq-5}
\ee
Looking at Fig.\,2 it is clear that the only singularities are found at $T=0$ (the system is finite) while all cusps and jumps are smoothened for $T>0$.

\begin{figure}
\begin{tabular}{ll}
{\rm a)}\qquad\qquad\qquad\qquad\,\,\,\,\,\,\,{\large\rm CT model} & {\rm b)}\qquad\qquad\qquad\qquad\,\,\,\,\,\,{\large\rm DC model}\\
\includegraphics[width=7.5 cm]{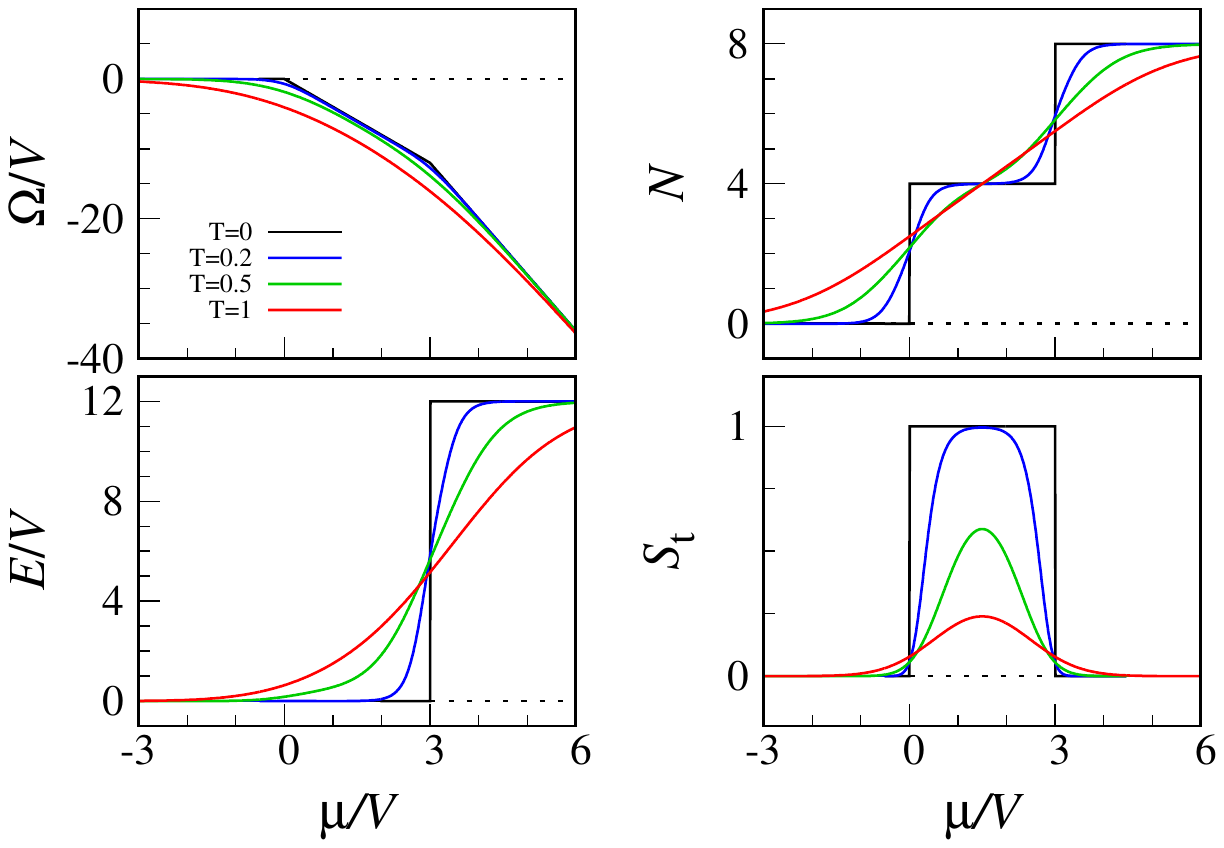} &
\includegraphics[width=7.5 cm]{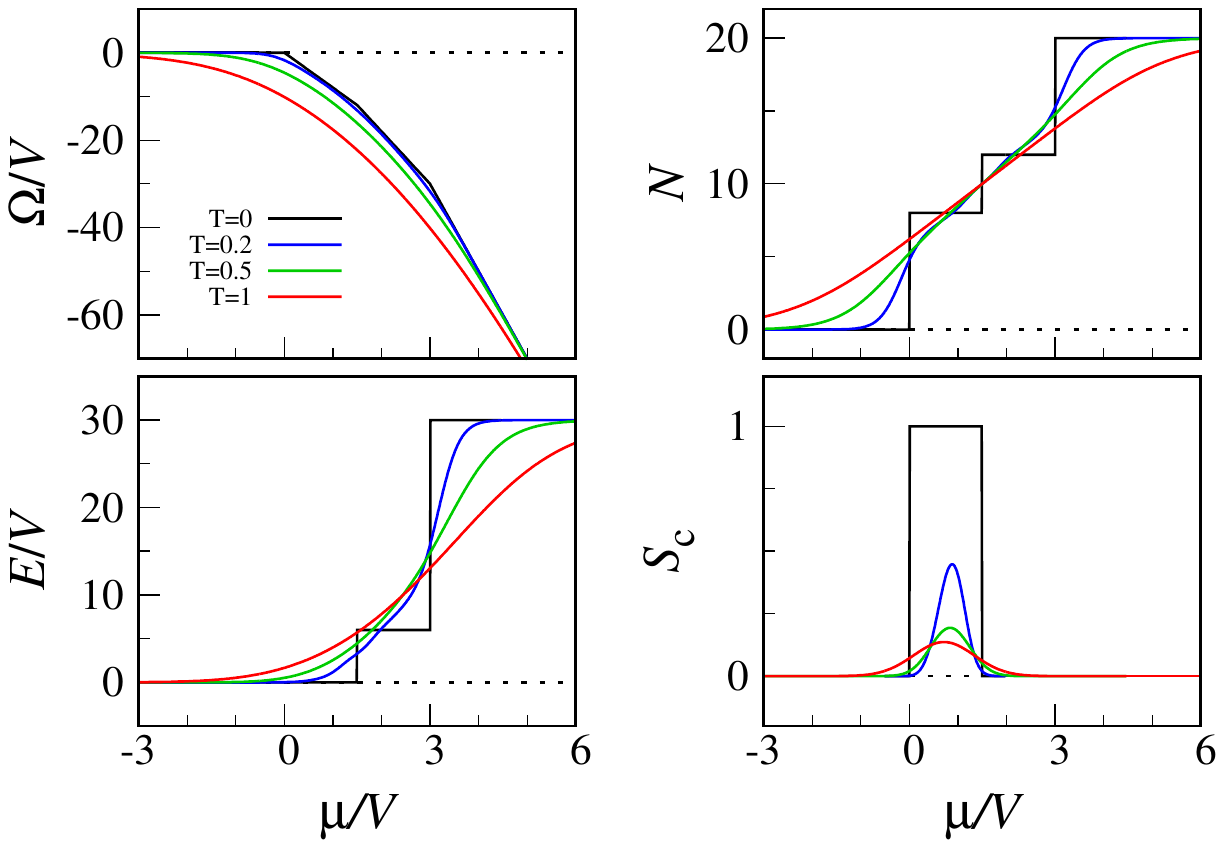}
\end{tabular}
\caption{Thermal averages for the CT model (a) and the DC model (b), plotted as a function of $\mu$ for four values of $T$ in units of $V/k_B$, where $k_B$ is Boltzmann's constant (see legend in the top left panel). Top left: grand potential. Top right: total number of occupied sites. Bottom left: total energy. Bottom right: order parameter (see text).}
\end{figure}

It is worth considering how the $T=0$ phase diagram of the CT and DC models gets modified when the occupancy of a node is allowed to take any value. For simplicity, as relevant $T=0$ states only the ``cluster crystals''~\cite{Likos,Mladek,Prestipino7} originated from the previously identified ground states are considered. Call $U>0$ the on-site energy and $V$ the NN repulsion. The grand Hamiltonian now reads $H[n]=U/2\sum_in_i(n_i-1)+V\sum_{\left<i,j\right>}n_in_j-\mu\sum_in_i$. For the CT model, the grand potential of a tetrahedral ``cluster crystal'' with $n$ particles per site is $\Omega_n^{(4)}=2n(n-1)U-4n\mu$, while the grand potential of the ``cluster crystal'' with $n$ particles per site is $\Omega_n^{(8)}=4n(n-1)U+12n^2V-8n\mu$. For a given $\mu$, the most stable ``phase'' corresponds to the minimum $\Omega$. Up to $\mu=0$ the stable ``phase'' is still the empty mesh. As $\mu$ grows further, the site occupancy increases monotonically within each of the families $\Omega_n^{(4)}$ and $\Omega_n^{(8)}$, but the exact sequence of stable ``phases'' depends on $U/V$. I show three cases in Fig.\,3. For $U=5V$ (right panel), the behavior for small $\mu>0$ recalls that found in the hard-core limit; however, as $\mu$ grows each site becomes increasingly populated, and a whole sequence of cluster states is found. The opposite occurs for $U=2V$ (left panel), where only the cluster states with checkerboard order are stabilized for $\mu>0$. A curious situation occurs for $U=3V$, where the competition between different cluster states becomes so stringent that ``crystals'' of the two families coexist at regular intervals along the $\mu$ axis (center panel).

\begin{figure}
\begin{tabular}{ccc}
\,\,\,$U=2V$ & \,\,\,$U=3V$ & \,\,\,$U=5V$\\
\includegraphics[width=4.85 cm]{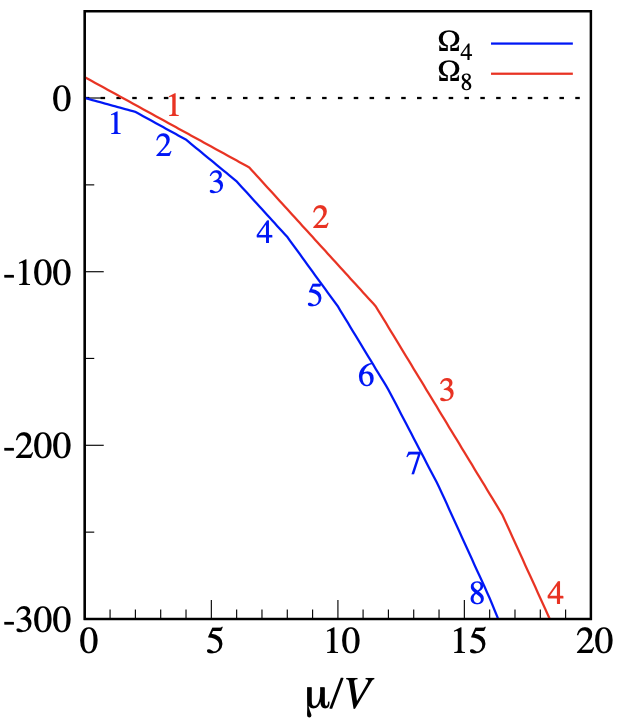} &
\includegraphics[width=4.85 cm]{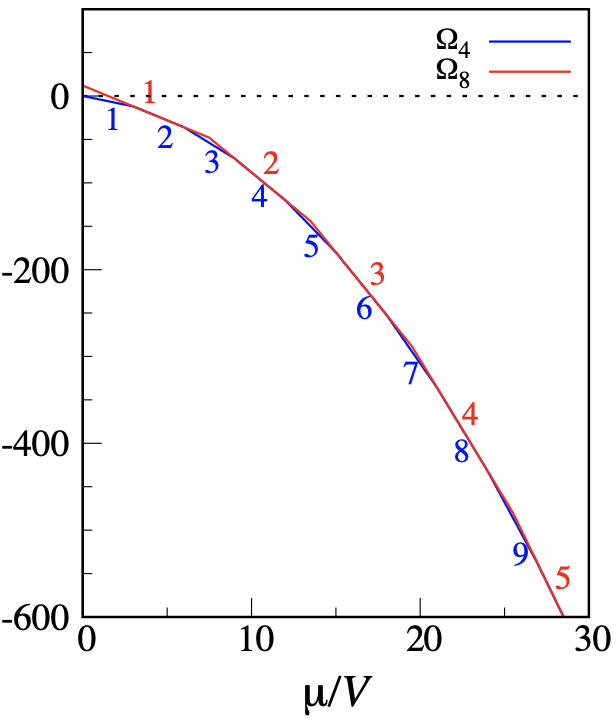} &
\includegraphics[width=4.8 cm]{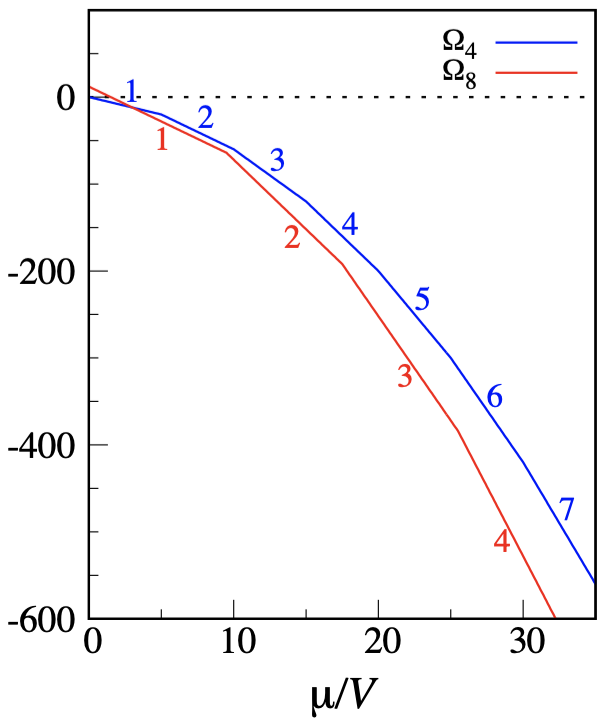}
\end{tabular}
\caption{CT model with multiple occupancy allowed. The grand potential $\Omega$ (in $V$ units) is plotted as a function of $\mu$ for two families of ``cluster crystals'', $\Omega_n^{(4)}$ (blue) and $\Omega_n^{(8)}$ (red), and for three values of $U/V$. Red and blue numbers are values of $n$.}
\end{figure}

As a matter of fact, hybrid ground states with features intermediate between those of the tetrahedral and cubic cluster states may also occur. For example, I have checked for $U=5V$ that a microstate with two overlapping particles at the vertices of one tetrahedron and one particle at the vertices of the other tetrahedron is the most stable ground state in the $\mu$ interval from $8V$ to $11V$.

Moving to the DC model, we have three different families of ``cluster phases'': the cubic family with grand potential $\Omega_n^{(8)}=4n(n-1)U-8n\mu$; the co-cubic family with grand potential $\Omega_n^{(12)}=6n(n-1)U+6n^2V-12n\mu$; and the family of those cluster states where every site is filled with the same number of particles, with grand potential $\Omega_n^{(20)}=10n(n-1)U+30n^2V-20n\mu$. Without delving into details, the phase behavior as a function of $U$ is similar to the CT model, with multiple coexistence between all three types of cluster crystals now occurring for $U=2V$.

\subsection{Quantum models}

The previous analysis has served to set the stage for the forthcoming study of the extended BH model on a regular spherical mesh; only the quantum versions of the CT and DC models (denoted QCT and QDC, respectively) are considered below.

The system is defined by the following grand Hamiltonian:
\be
H=-t\sum_{\left<i,j\right>}\left(a_i^\dagger a_j+a_j^\dagger a_i\right)+\frac{U}{2}\sum_in_i(n_i-1)+V\sum_{\left<i,j\right>}n_in_j-\mu\sum_in_i\,.
\label{eq-6}
\ee
In the present investigation, $H$ describes a system of bosons on a spherical mesh of $M$ sites, $a_i,a_i^\dagger$ are bosonic field operators ($i=1,\ldots,M$), and $n_i=a_i^\dagger a_i$ is a number operator. Moreover, $t\ge 0$ is the hopping amplitude between NN sites, $U>0$ is the on-site repulsion, and $V>0$ is the strength of the NN repulsion. In the hard-core limit $U\rightarrow\infty$, the site occupancy is restricted to zero or one and the $U$ term in (\ref{eq-6}) vanishes; it is only this limit that is treated hereafter.

In the BH model on a standard lattice, at $T=0$ we observe an insulator-superfluid transition with increasing $t$ for every positive $\mu$. The addition of $V$ may stabilize (depending on the lattice) a density wave at low $t$, as well as a supersolid phase at the boundary between the insulator and superfluid phases. These features are also present in the hard-core limit, as reported, e.g., in Refs.~\cite{Zhang,Gheeraert}.

\section{MF investigation}

MF theory is the method of choice when a new many-body problem is attacked; it has been frequently applied for continuous quantum systems as well, as an effective means to identify the ground states and quantum transitions between them (cf. Refs.~\cite{Kunimi,Macri,Prestipino8}). Various versions of MF theory exist in discrete space; here the so-called {\em decoupling approximation}~\cite{Sheshadri,Kurdestany,Gheeraert} is employed, which gives the advantage of a fully analytic treatment of the problem. Clearly, the accuracy of MF theory would be questionable when applied for a finite quantum system, even one lacking a boundary surface. This will make urgent an assessment of MF theory against exact results, which however is delayed until the next Section.

In the decoupling approximation, the two-site terms in the Hamiltonian (\ref{eq-6}) are linearized so that the eigenvalue problem becomes effectively one-site. This is accomplished by first writing the exact identity
\be
a_i^\dagger a_j=a_i^\dagger\left<a_j\right>+\left<a_i^\dagger\right>a_j-\left<a_i^\dagger\right>\left<a_j\right>+\delta a_i^\dagger\delta a_j
\label{eq-7}
\ee
with $\delta a_i=a_i-\left<a_i\right>$ (for $i=1,\ldots,M$), and then neglecting the last term. Similarly,
\be
n_in_j\approx n_i\left<n_j\right>+\left<n_i\right>n_j-\left<n_i\right>\left<n_j\right>\,.
\label{eq-8}
\ee
The averages $\left<a_i\right>\equiv\phi_i$ and $\left<n_i\right>\equiv\rho_i$ (either ground-state averages  or thermal averages) are to be determined self-consistently; $\phi_i$ and $\rho_i$ (with $0\le\rho_i\le 1$) represent the superfluid order parameter and the local density for site $i$, respectively. For hard-core bosons we readily obtain $H\approx H_{\rm MF}$ with
\be
H_{\rm MF}=-t\sum_i\left(F_ia_i^\dagger+F_i^*a_i-F_i\phi_i^*\right)+\frac{V}{2}\sum_i\left(2R_in_i-R_i\rho_i\right)-\mu\sum_in_i\,.
\label{eq-9}
\ee
In Eq.\,(\ref{eq-9}), $F_i=\sum_{j\in{\rm NN}_i}\phi_j$ and $R_i=\sum_{j\in{\rm NN}_i}\rho_j$ are sums over the nearest neighbors of the $i$-th site. For a bipartite mesh, sites are either of type $A$ or $B$, hence the unknown parameters are four, namely $\phi_A,\phi_B,\rho_A,\rho_B$. The simplification is obvious: rather than working on a Fock space of dimensionality $2^M$, for a bipartite mesh the basis states are just four, namely $\left|0,0\right>,\left|0,1\right>,\left|1,0\right>$, and $\left|1,1\right>$, corresponding to the possible occupancies of a pair of A and B sites.

\subsection{QCT model}

For the QCT model, the mesh is bipartite and formed by two tetrahedral sub-meshes with four points each. A point of grid $A$ has three neighbors, all belonging to grid $B$, and {\em vice versa}. Hence:
\be
F_A=3\phi_B\,,\,\,\,F_B=3\phi_A\,,\,\,\,R_A=3\rho_B\,,\,\,\,{\rm and}\,\,\,R_B=3\rho_A\,.
\label{eq-10}
\ee
Using a subscript $A$ ($B$) for operators relative to a single $A$ ($B$) site, the MF Hamiltonian reads:
\be
H_{\rm MF}=E_0-12t\left(\phi_Ba_A^\dagger+\phi_B^*a_A+\phi_Aa_B^\dagger+\phi_A^*a_B\right)+4(3V\rho_B-\mu)n_A+4(3V\rho_A-\mu)n_B
\label{eq-11}
\ee
with
\be
E_0=12t(\phi_A\phi_B^*+\phi_A^*\phi_B)-12V\rho_A\rho_B\,.
\label{eq-12}
\ee
On the 4-vector basis ${\cal B}=\left\{\left|0,0\right>,\left|0,1\right>,\left|1,0\right>,\left|1,1\right>\right\}$ the Hamiltonian (\ref{eq-11}) is represented by a $4\times 4$ Hermitian matrix:
\be
H_{\rm MF}=\begin{pmatrix}
E_0 & -12t\phi_A^* & -12t\phi_B^* & 0\\
-12t\phi_A & E_0+4(3V\rho_A-\mu) & 0 & -12t\phi_B^*\\
-12t\phi_B & 0 & E_0+4(3V\rho_B-\mu) & -12t\phi_A^*\\
0 & -12t\phi_B & -12t\phi_A & E_0+12V(\rho_A+\rho_B)-8\mu
\end{pmatrix}\,.
\label{eq-13}
\ee
Non-superfluid phases have $\phi_A=\phi_B=0$. In this case the matrix becomes diagonal, meaning that the basis vectors are all (grand-)energy eigenstates. From the self-consistency equations $\rho_A=\left<n_A\right>$ and $\rho_B=\left<n_B\right>$, it readily follows that $\rho_A=\rho_B=0$ for $\left|0,0\right>$ (empty mesh), $\rho_A=0$ and $\rho_B=1$ for $\left|0,1\right>$ (tetrahedral-$B$ ``crystal''), $\rho_A=1$ and $\rho_B=0$ for $\left|1,0\right>$ (tetrahedral-$A$ ``crystal''), and $\rho_A=\rho_B=1$ for $\left|1,1\right>$ (fully occupied mesh). The eigenvalue gives the grand potential $\Omega$, which is zero for the empty mesh, $-4\mu$ for the twofold-degenerate tetrahedral ``crystal'', and $12V-8\mu$ for the filled mesh. These grand potentials are exactly the same as in the CT model.

Superfluid and supersolid ``phases'' have non-zero, possibly distinct complex values of $\phi_A$ and $\phi_B$. For sure, $\phi_A$ and $\phi_B$ have equal phases since only the magnitude of the order parameter can be spatially modulated, hence the arbitrary phase can be taken as zero. Without loss of generality, we may assume $\phi_A$ and $\phi_B$ to be positive quantities. We shall see in the next Section that the exact eigenstates of $H$ for $t>0$ do not distinguish between $A$ and $B$, hence our search can be restricted to homogeneous (superfluid) solutions: $\phi_A=\phi_B=\phi$ and $\rho_A=\rho_B=\rho$. We are thus led to the following characteristic equation (with $E_0=24t\phi^2-12V\rho^2$):
\be
\begin{vmatrix}
E_0-\lambda & -12t\phi & -12t\phi & 0\\
-12t\phi & E_0+4(3V\rho-\mu)-\lambda & 0 & -12t\phi\\
-12t\phi & 0 & E_0+4(3V\rho-\mu)-\lambda & -12t\phi\\
0 & -12t\phi & -12t\phi & E_0+8(3V\rho-\mu)-\lambda
\end{vmatrix}
=0\,.
\label{eq-14}
\ee
Doing the simplifications, we arrive at
\be
(\lambda-b)^2\left[(\lambda-a)(\lambda-2b+a)-4u^2\right]=0
\label{eq-15}
\ee
with $a=E_0,b=E_0+4(3V\rho-\mu)$, and $u=-12t\phi$. The minimum root of (\ref{eq-15}) is
\be
E=b-\sqrt{(a-b)^2+4u^2}=4\left(6t\phi^2-3V\rho^2+3V\rho-\mu-\sqrt{(3V\rho-\mu)^2+36t^2\phi^2}\right)\,.
\label{eq-16}
\ee
A real eigenvector of $E$ is
\be
\left|\psi_E\right>=\frac{E-2b+a}{2u}\left|0,0\right>+\left|0,1\right>+\left|1,0\right>+\frac{E-a}{2u}\left|1,1\right>\,,
\label{eq-17}
\ee
or, in explicit terms,
\be
\left|\psi_E\right>=\left(\frac{3V\rho-\mu+\sqrt{(3V\rho-\mu)^2+36t^2\phi^2}}{6t\phi},1,1,\frac{-(3V\rho-\mu)+\sqrt{(3V\rho-\mu)^2+36t^2\phi^2}}{6t\phi}\right)\,.
\label{eq-18}
\ee
Now imposing the conditions
\be
\rho_{A,B}=\frac{\left<\psi_E|n_{A,B}|\psi_E\right>}{\left<\psi_E|\psi_E\right>}\,\,\,\,\,\,{\rm and}\,\,\,\,\,\,\phi_{A,B}=\frac{\left<\psi_E|a_{A,B}|\psi_E\right>}{\left<\psi_E|\psi_E\right>}\,,
\label{eq-19}
\ee
we arrive at the two coupled equations
\be
1-2\rho=\frac{3V\rho-\mu}{\sqrt{(3V\rho-\mu)^2+36t^2\phi^2}}\,\,\,\,\,\,{\rm and}\,\,\,\,\,\,1=\frac{3t}{\sqrt{(3V\rho-\mu)^2+36t^2\phi^2}}\,,
\label{eq-20}
\ee
which are easily solved to give
\be
\rho=\frac{\mu+3t}{3V+6t}\,\,\,\,\,\,{\rm and}\,\,\,\,\,\,\phi=\frac{\sqrt{(\mu+3t)(3V+3t-\mu)}}{3V+6t}\,.
\label{eq-21}
\ee
The above $\phi$ solution only exists provided that $-3t\le\mu\le 3V+3t$, which is a necessary condition for the existence of the superfluid. Plugging these $\rho$ and $\phi$ in Eq.~(\ref{eq-16}), we finally obtain the superfluid grand potential:
\be
\Omega_{\rm SF}=-\frac{4(\mu+3t)^2}{3V+6t}\,.
\label{eq-22}
\ee
This outcome can also be obtained from the general relation (see Eqs.\,(\ref{eq-6})-(\ref{eq-8}))
\be
\left<H_{\rm MF}\right>=-2t\sum_{\left<i,j\right>}\phi_i^*\phi_j+V\sum_{\left<i,j\right>}\rho_i\rho_j-\mu\sum_i\rho_i\,.
\label{eq-23}
\ee
For the superfluid we have $\phi_i=\phi$ and $\rho_i=\rho$, with $\phi$ and $\rho$ given by Eq.~(\ref{eq-21}), hence:
\be
\left<H_{\rm MF}\right>=4\left(-6t\phi^2+3V\rho^2-2\mu\rho\right)=-\frac{4(\mu+3t)^2}{3V+6t}\,.
\label{eq-24}
\ee

By comparing the grand potentials of all the ``phases'' we arrive at the ground-state diagram in Fig.\,4. Along the straight lines $\mu=-3t$ and $\mu=3V+3t$, separating the superfluid from the insulator ``phases'' at small and large $\mu$, the $\mu$ derivative of the grand potential ($=-\left<N\right>$) is continuous. Instead, along the line
\be
\mu=\frac{3V}{2}\pm\frac{3}{2}\sqrt{V^2-4t^2}\,,
\label{eq-25}
\ee
separating the ``crystalline'' lobe from the superfluid, the average number of occupied sites shows a jump discontinuity. Only at the lobe vertex $(V/2,3V/2)$ the tetrahedral-superfluid transition is continuous. We will see in Section 4 to what extent the indications of MF theory are accurate, considering that the cubic mesh consists of 8 sites only.

\begin{figure}
\centering
\includegraphics[width=10 cm,angle=-90]{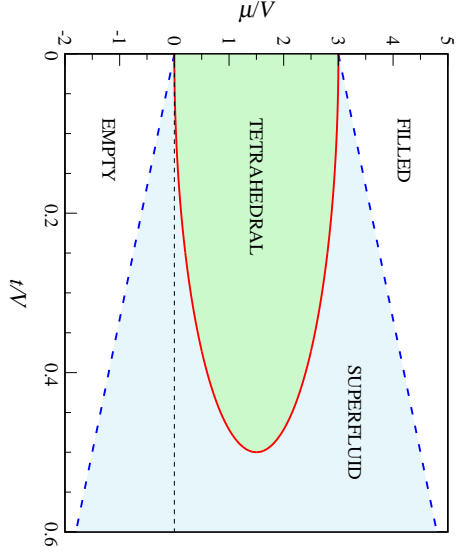}
\caption{Phase diagram of the QCT model according to the decoupling approximation (see text). The blue dashed lines are second-order transition lines, while the red full line is a first-order line (only at the lobe vertex the transition is continuous).}
\end{figure}

\subsection{QDC model}

The decoupling approximation for the QDC model works similarly as for the QCT model. Again, the starting point is Eq.~(\ref{eq-9}) and the mesh is partitioned into a cube ($A$, 8 sites) and a co-cube ($B$, 12 sites). We have:
\be
F_A=3\phi_B\,,\,\,\,F_B=2\phi_A+\phi_B\,,\,\,\,R_A=3\rho_B\,,\,\,\,{\rm and}\,\,\,R_B=2\rho_A+\rho_B\,.
\label{eq-26}
\ee
The MF Hamiltonian then reads:
\ba
H_{\rm MF}&=&E_0-24t(\phi_Ba_A^\dagger+\phi_B^*a_A)-12t\left((2\phi_A+\phi_B)a_B^\dagger+(2\phi_A^*+\phi_B^*)a_B\right)
\nonumber \\
&+&4(6V\rho_B-2\mu)n_A+4(6V\rho_A+3V\rho_B-3\mu)n_B
\label{eq-27}
\ea
with
\be
E_0=24t(\phi_A\phi_B^*+\phi_A^*\phi_B)+12t|\phi_B|^2-24V\rho_A\rho_B-6V\rho_B^2\,.
\label{eq-28}
\ee
On the ${\cal B}$ basis the Hamiltonian (\ref{eq-27}) is represented by the matrix:
\small
\be
H_{\rm MF}=\begin{pmatrix}
E_0 & -12t(2\phi_A^*+\phi_B^*) & -24t\phi_B^* & 0\\
-12t(2\phi_A+\phi_B) & E_0+4(6V\rho_A+3V\rho_B-3\mu) & 0 & -24t\phi_B^*\\
-24t\phi_B & 0 & E_0+4(6V\rho_B-2\mu) & -12t(2\phi_A^*+\phi_B^*)\\
0 & -24t\phi_B & -12t(2\phi_A+\phi_B) & E_0+4(6V\rho_A+9V\rho_B-5\mu)
\end{pmatrix}\,.
\label{eq-29}
\ee
\normalsize
For phases with $\phi_A=\phi_B=0$ the matrix is diagonal and, like in the QCT model, the basis vectors are all (grand-)energy eigenstates. From the self-consistency conditions $\rho_A=\left<n_A\right>$ and $\rho_B=\left<n_B\right>$ it follows that $\rho_A=\rho_B=0$ for $\left|0,0\right>$ (empty mesh), $\rho_A=0$ and $\rho_B=1$ for $\left|0,1\right>$ (co-cubic ``crystal''), $\rho_A=1$ and $\rho_B=0$ for $\left|1,0\right>$ (cubic ``crystal''), and $\rho_A=\rho_B=1$ for $\left|1,1\right>$ (filled mesh). The grand potential $\Omega$ is zero for the vacuum, $-8\mu$ for the cubic ``crystal'', $6V-12\mu$ for the co-cubic ``crystal'', and $30V-20\mu$ for the filled mesh. These values are the same as for the DC model, hence the same sequence of ``phases'' as a function of $\mu$ is observed in the QDC model for $t=0$.

Moving to phases with $\phi_A,\phi_B\ne 0$, I first consider the possibility of a superfluid ($\phi_A=\phi_B=\phi>0$ and $\rho_A=\rho_B=\rho$). In this case the characteristic equation takes the form:
\be
\begin{vmatrix}
a-\lambda & u & v & 0\\
u & b-\lambda & 0 & v\\
v & 0 & c-\lambda & u\\
0 & v & u & d-\lambda
\end{vmatrix}
=0
\label{eq-30}
\ee
with $E_0=60t\phi^2-30V\rho^2,a=E_0,b=E_0+12(3V\rho-\mu),c=E_0+8(3V\rho-\mu),d=E_0+20(3V\rho-\mu),u=-36t\phi,v=-24t\phi$. Observing that $a+d=b+c\equiv s$, the minimum eigenvalue of $H_{\rm MF}$ turns out to be (see Appendix A):
\ba
E&=&\frac{1}{2}\left(s-\sqrt{(a-b)^2+4u^2}-\sqrt{(a+b-s)^2+4v^2}\right)
\nonumber \\
&=&10\left(6t\phi^2-3V\rho^2+3V\rho-\mu-\sqrt{(3V\rho-\mu)^2+36t^2\phi^2}\right)\,.
\label{eq-31}
\ea
Notice that this energy is exactly $5/2$ times that of the QCT model (see Eq.\,(\ref{eq-16})).

Using $p=(b-c)/2=(d-a)/10=2(3V\rho-\mu)$ and $q=u/3=v/2=-12t\phi$, the coordinates $(x,y,z,w)$ of an eigenvector $\left|\psi_E\right>$ of $E$ satisfy the following linear system:
\be
\left\{
\begin{array}{ll}
-5px+3qy+2qz=-5\sqrt{p^2+q^2}\,x & \\
3qx+py+2qw=-5\sqrt{p^2+q^2}\,y & \\
2qx-pz+3qw=-5\sqrt{p^2+q^2}\,z & \\
2qy+3qz+5pw=-5\sqrt{p^2+q^2}\,w\,,
\end{array}
\right.
\label{eq-32}
\ee
implying that $\left|\psi_E\right>$ is also eigenvector of the simpler matrix
\be
\begin{pmatrix}
-5p & 3q & 2q & 0\\
3q & p & 0 & 2q\\
2q & 0 & -p & 3q\\
0 & 2q & 3q & 5p
\end{pmatrix}
\label{eq-33}
\ee
with eigenvalue $-5\sqrt{p^2+q^2}$. For $t>0$, one such vector is
\be
\left|\psi_E\right>=\left(-\frac{p+\sqrt{p^2+q^2}}{q},1,1,\frac{p-\sqrt{p^2+q^2}}{q}\right)\,,
\label{eq-34}
\ee
or
\be
\left|\psi_E\right>=\left(\frac{3V\rho-\mu+\sqrt{(3V\rho-\mu)^2+36t^2\phi^2}}{6t\phi},1,1,\frac{-(3V\rho-\mu)+\sqrt{(3V\rho-\mu)^2+36t^2\phi^2}}{6t\phi}\right)\,.
\label{eq-35}
\ee
This is identical to the state (\ref{eq-18}) describing the superfluid in the QCT model. The reason of this equivalence is that, in both cubic and dodecahedral meshes, every site has three neighbors; hence, in the superfluid the quantities $F_A,F_B,R_A,R_B$ are the same. No surprise, then, if also the expressions of $\rho$ and $\phi$ turn out to be equal for the two models (the consistency equations are identical). As already commented, the energy of the QDC superfluid is instead a factor $5/2$ larger (in absolute terms) than in the QCT model:
\be
\Omega_{\rm SF}=-\frac{10(\mu+3t)^2}{3V+6t}\,.
\label{eq-36}
\ee

The main novelty with respect to the QCT model is the existence of a stable supersolid ``phase'' in the QDC model. To seek for MF solutions having the character of a supersolid, I have first derived the exact equations for the MF parameters $\rho_A,\rho_B,\phi_A,\phi_B$, without {\em a priori} assuming them to be equal in pairs (this is done in Appendix B). Then, I have solved these equations numerically so as to find the region where the supersolid is more stable than the other phases. This task is made simpler by noting that, thanks to a symmetry property of the equations, from a supersolid solution with $\mu>3V/2$ it is possible to obtain another solution with $\mu<3V/2$. Indeed, it easily follows from Eqs.\,(\ref{b-6}) that, if $(\rho_A,\rho_B,\phi_A,\phi_B)$ is a solution for a certain $\mu$, then $(1-\rho_A,1-\rho_B,\phi_A,\phi_B)$ is a solution for $3V-\mu$. Moreover, for $\mu=3V/2$ the two solutions share the same grand potential. It turns out that the $(t,\mu)$ region where the supersolid is stable is symmetric about the $\mu=3V/2$ axis, lying across the boundary between the solid ``phases'' and the superfluid (see Fig.\,5). To all evidence, the boundary between the supersolid and the superfluid lies at $t=1/3$. In fact, the supersolid consists of two distinct ``phases'', SS1 below $\mu=3V/2$ (where $\rho_B<1/2<\rho_A$) and SS2 above $\mu=3V/2$ (where $\rho_A<1/2<\rho_B$), with $\phi_A<\phi_B<1/2$ in both. In SS1 cubic sites are more occupied, on average, than co-cubic sites, whereas the opposite occurs in SS2. The two supersolids coexist for $\mu=3V/2$, for all $t$ in the range 0.25 to $1/3$. Hard-core bosons on the triangular lattice have a similar phase diagram~\cite{Gheeraert}, but in that case the supersolid extends down to $t=0$ and the solid lobes do not overlap each other.

\begin{figure}
\centering
\includegraphics[width=10 cm,angle=-90]{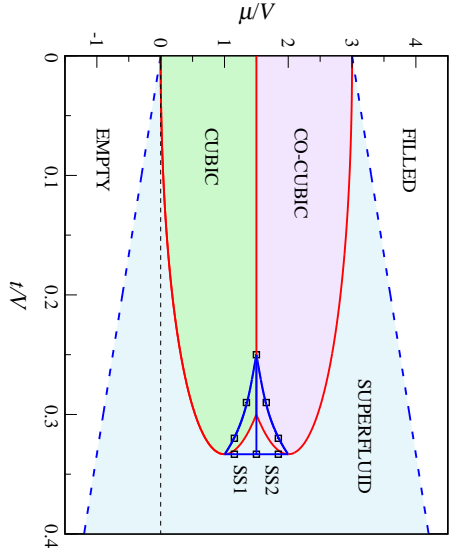}
\caption{Phase diagram of the QDC model according to the decoupling approximation (see text). The blue dashed lines are second-order transition lines, while red full lines are first-order lines (only at the lobe vertices the transition is continuous). Inside the blue circuit, the stable phase is supersolid (SS1 below $\mu=3V/2$; SS2 above $\mu=3V/2$). The full lines through the data points (open black squares) are a guide for the eye.}
\end{figure}

The full $T=0$ phase diagram of the QDC model according to the decoupling approximation is reported in Fig.\,5. Along the straight lines $\mu=-3t$ and $\mu=3V+3t$, the $\mu$ derivative of the grand potential is continuous. Instead, along the curves
\be
\mu=\frac{-3t+6V\pm 6\sqrt{V^2-Vt-6t^2}}{5}\,\,\,\,\,\,{\rm and}\,\,\,\,\,\,\mu=\frac{3t+9V\pm 6\sqrt{V^2-Vt-6t^2}}{5}\,,
\label{eq-37}
\ee
which respectively separate the cubic and co-cubic regions from the superfluid region, the total number of particles jumps discontinuously. Only at the lobe vertices $(V/3,V)$ and $(V/3,2V)$ the two ``crystal''-superfluid transitions are continuous. The line $\mu=3V/2$ separating the two ``crystalline'' regions ends at the point $(3V/10,3V/2)$ where the two boundary curves (\ref{eq-37}) cross each other. However, this triple point is only metastable since, in the whole region bounded by the blue circuit of Fig.\,5, the stable phase is actually supersolid.

\section{Assessment of MF theory}

Let us again reconsider the QCT model at $T=0$. Due to the relatively small dimensionality of its Hilbert space ($=256$), the model can be solved numerically, determining (among others) the exact ground state $\left|g\right>$ and its eigenvalue (i.e., the grand potential) as a function of $t$ and $\mu$. This is obtained by representing the grand Hamiltonian $H$ on the Fock basis $\{\left|x_1,x_2,\ldots,x_8\right>\}$ (with $x_i=0$ or 1 for all $i$) and diagonalizing the ensuing matrix. An extensive mapping of a few quantum averages enable us to clarify the nature of the ``phases'' present.

As usual, the sites of the mesh are classified according to what tetrahedral sub-mesh, $A$ or $B$, they belong to. Then, I compute the average occupancies of $A$ and $B$ sites (corresponding to the MF parameters $\rho_A$ and $\rho_B$, respectively), the average values of $a_A$ and $a_B$ (corresponding to the MF parameters $\phi_A$ and $\phi_B$, respectively), and the {\em superfluid density} $\rho_{\rm SF}$ (see, e.g., Ref.~\cite{vanOosten,Yamamoto}). In the present case, the latter quantity reads:
\be
\rho_{\rm SF}\equiv\frac{1}{8}\left<g\left|\widetilde{a}_{\bf 0}^\dagger\widetilde{a}_{\bf 0}\right|g\right>=\frac{1}{64}\sum_{i,j=1}^8\left<g\left|a_i^\dagger a_j\right|g\right>=\frac{1}{8}\sum_{j=1}^8\left<g\left|a_1^\dagger a_j\right|g\right>\,,
\label{eq-38}
\ee
where $\widetilde{a}_{\bf 0}=(1/\sqrt{8})\sum_{i=1}^8a_i$ is the zero-momentum field operator. Notice that, in a large lattice of $M$ sites, $\left<\widetilde{a}_{\bf 0}^\dagger\widetilde{a}_{\bf 0}\right>=N_0$ is the average number of condensate particles, hence $\rho_{\rm SF}=N_0/M$ is indeed the condensate density.

As far as the occupancies are concerned, exact diagonalization shows that they are always equal for an $A$ and a $B$ site, with the only exception of $t=0$ and $0<\mu<3V$ where the occupancies are as in MF theory (i.e., either $\left<n_A\right>=1$ and $\left<n_B\right>=0$ or {\em vice versa}). In particular, for $t>0$ the equivalence $\left<n_A\right>=\left<n_B\right>$ holds in the whole $(t,\mu)$ region pertaining, according to MF theory, to the tetrahedral ``phase''. Hence, no spontaneous symmetry breaking does really occur in the QCT model, except for $t=0$. Indeed, for $t>0$ the Fourier coefficients of $\left|g\right>$ relative to any pair of Fock states equal by $A$-$B$ inversion are the same. I show in Fig.\,6 the average site occupancy in the whole space of parameters. We see that the $(t,\mu)$ plane is divided in zones where the site occupancy, which overall grows with $\mu$, takes the same constant value. The possible values are $k/8$, for $k=0,1,\ldots,8$. This outcome is not entirely unexpected considering that the Hamiltonian commutes with the $N$ operator, implying that the ground state (which is non-degenerate for $t>0$) should also be an eigenstate of $N$. Like in MF theory, the occupancy is zero below $\mu=-3t$ and 1 above $\mu=3V+3t$; in a whole region around $\mu=3V/2$ the occupancy is 0.5 for all $t$, with no abrupt transition from ``crystal'' to superfluid values.

\begin{figure}
\centering
\includegraphics[width=13 cm]{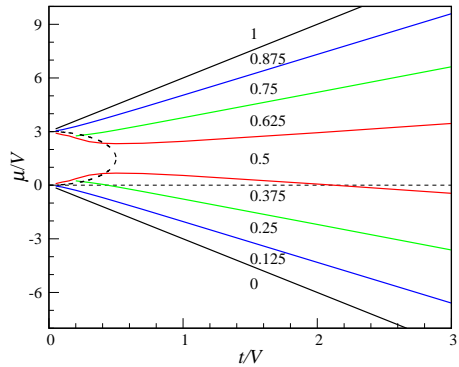}
\caption{Exact average site occupancy in the QCT model (full symmetry occurs between the $A$ and $B$ sub-meshes). The plotted numbers are the occupancies in the whole regions delimited by the coloured lines. These values reflect a perfect particle-hole symmetry around $\mu=3V/2$. The MF boundary between the tetrahedral and superfluid ``phases'' is also shown for comparison (dashed line).}
\end{figure}

Another difference with MF theory concerns the ground-state averages of $a_A$ and $a_B$: these are identically zero for all $t$ and $\mu$ values, which may seem in stark contrast to the behavior of the MF parameter $\phi$. In fact, it is not these averages that should be monitored in an exact treatment but rather the $\rho_{\rm SF}$ quantity defined at Eq.~(\ref{eq-38}). In the top panel of Fig.\,7, I have reported the $\rho_{\rm SF}$ values computed along a few iso-$t$ lines; each jump discontinuity of $\rho_{\rm SF}$ occurs in coincidence with one of the site occupancy. For comparison, in the bottom panel of Fig.\,7 I show the MF values of $\phi^2$ along the same lines. We see a clear correlation between the two behaviors, which demonstrates the existence of signatures of superfluidity in a finite quantum system. However, intriguing differences also exist: i) first observe that $\rho_{\rm SF}=0.125$ in the filled mesh, which is an oddity for a perfectly insulating state. In fact, 0.125 is $1/M$ for $M=8$, suggesting that this is an artifact of the finite system size. ii) Another finite-size effect are the non-zero values of $\rho_{\rm SF}$ in the purported tetrahedral region. However, $\rho_{\rm SF}$ is here sufficiently small that its peaks in the $\mu$ gaps between the tetrahedral and insulator regions remain well visible --- in MF theory these gaps fall in the superfluid region.

\begin{figure}
\centering
\includegraphics[width=8 cm,angle=-90]{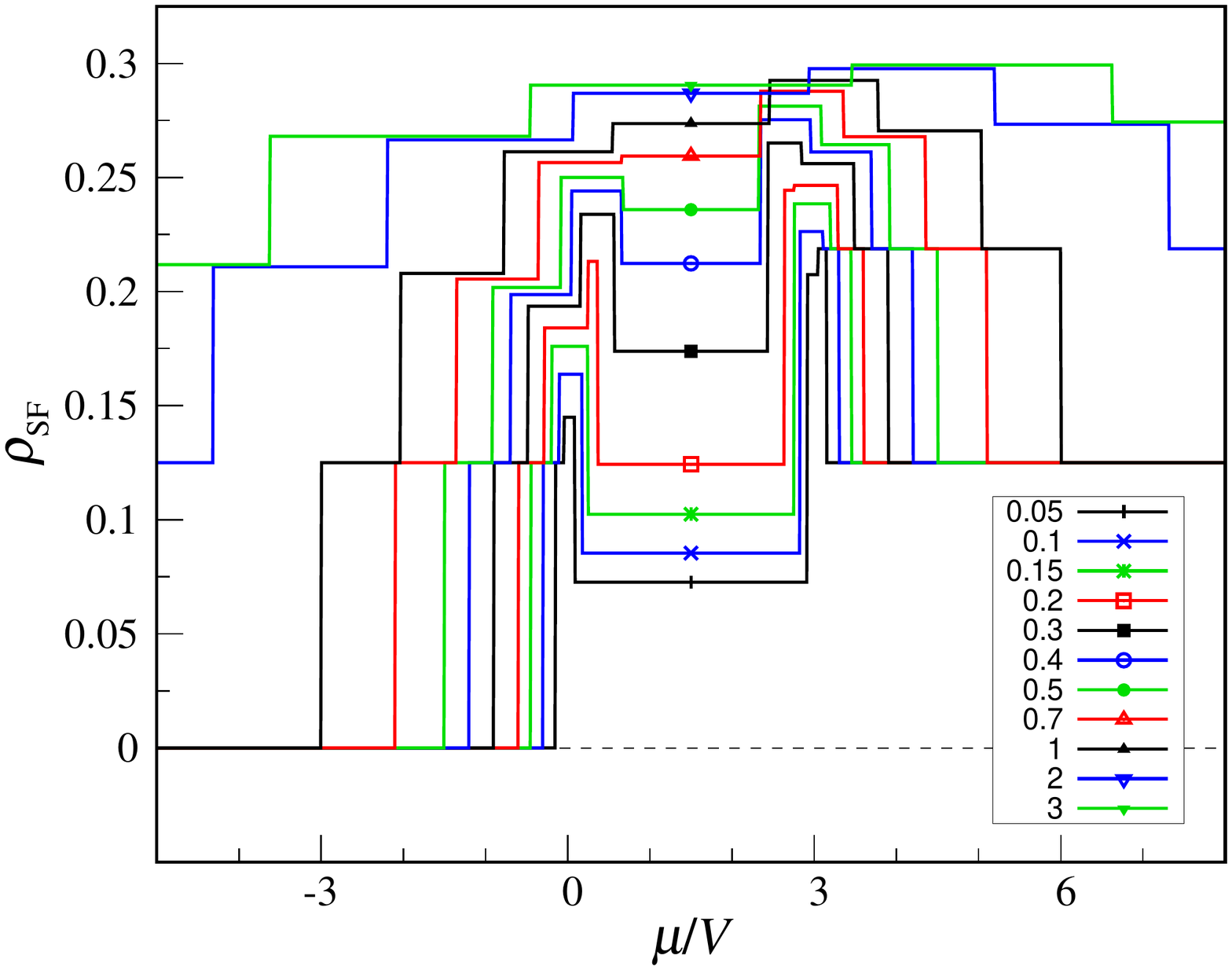}
\includegraphics[width=8 cm,angle=-90]{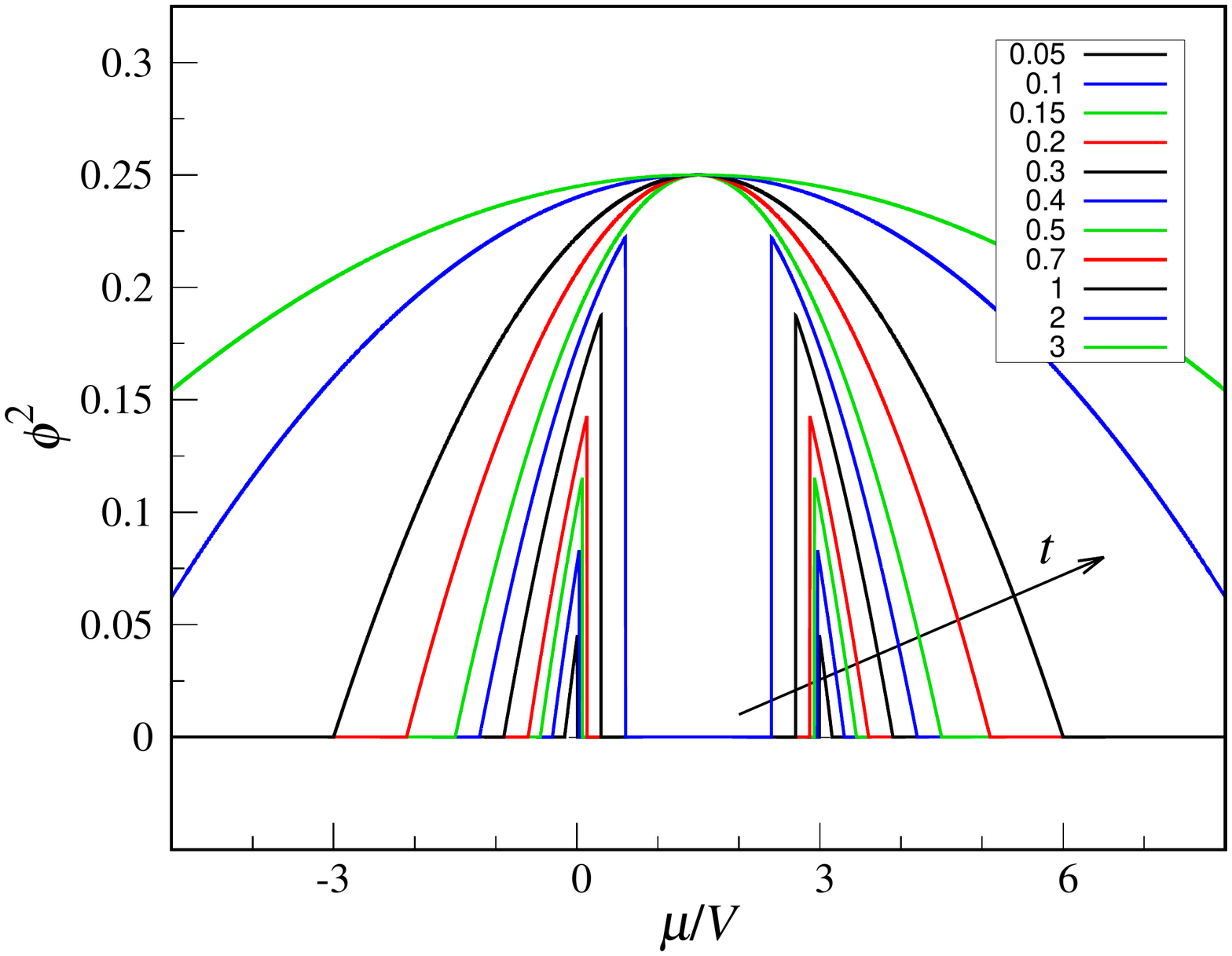}
\caption{Superfluid density in the QCT model: comparison between exact diagonalization (top panel) and decoupling approximation (bottom panel). The superfluid density is plotted as a function of $\mu$ for a number of $t$ values in units of $V$, increasing from bottom to top (see legends).}
\end{figure}

Finally, in Fig.\,8 I plot the QCT grand potential as a function of $\mu$ for a number of $t$ values in the range 0 to 0.5. In the MF curves, cusp singularities are associated with the crossing of first-order transition lines. We see that MF theory systematically overestimates exact values, the more so the larger $t$ is.

\begin{figure}
\centering
\includegraphics[width=8 cm,angle=-90]{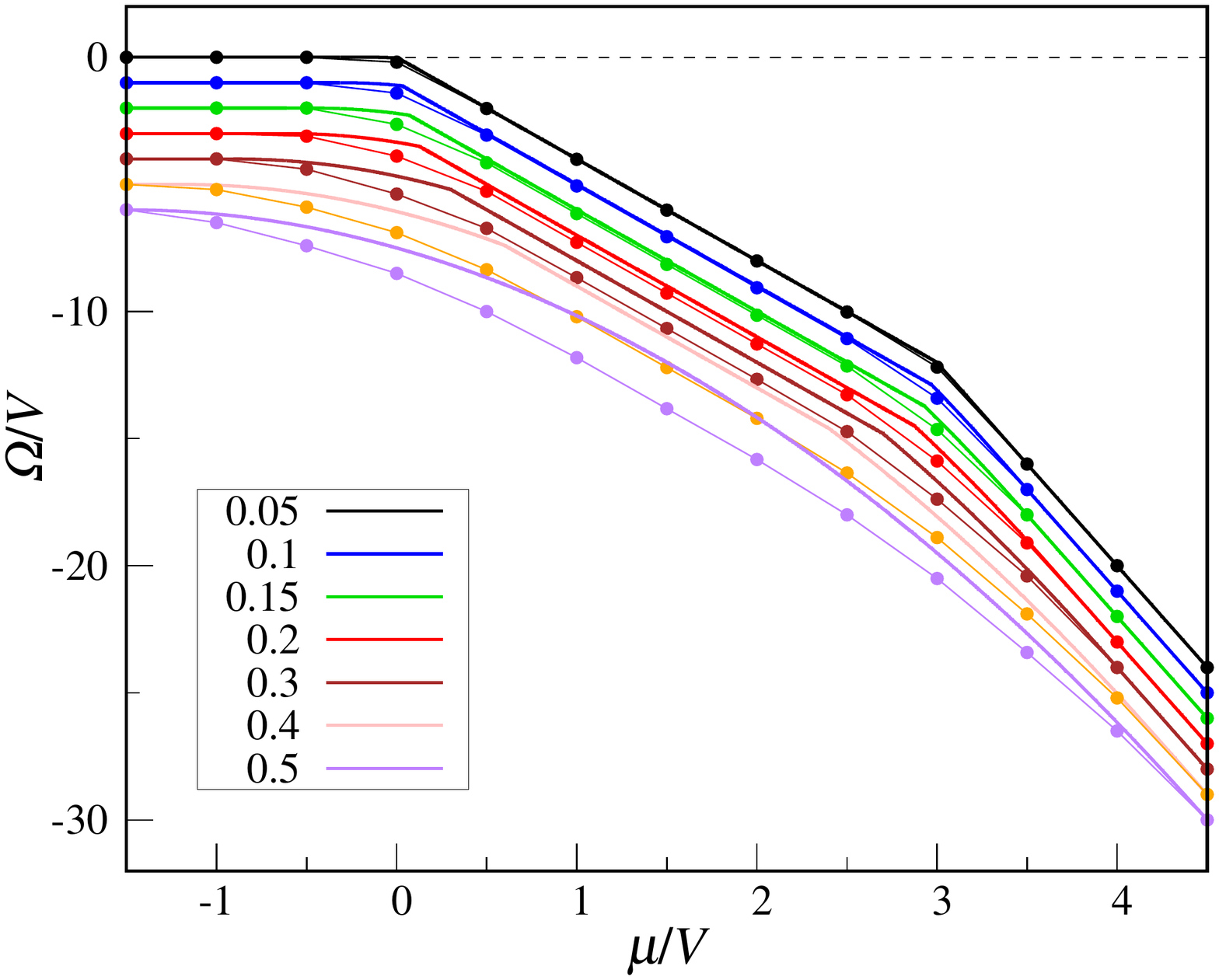}
\caption{Grand potential of the QCT model as a function of $\mu$, for a number of $t$ values (from top to bottom, $t=0.05,0.1,0.15,0.2,0.3,0.4,0.5$). MF data (thick full lines) are compared with exact values (full dots). To improve figure readability, the dots are joined by thin straight-line segments and the data are displaced vertically by 1 with respect to one another, starting from $t=0.1$.}
\end{figure}

\section{Conclusions}

I have worked out the zero-temperature phase diagram of two systems of hard-core bosons defined on the nodes of a regular spherical mesh. The interaction is of Bose-Hubbard type, with a further repulsion between neighboring particles. Choosing a suitable mesh, bosons are pushed to form a Platonic ``crystal'' in a range of chemical potentials: in the QCT model, the mesh is cubic and a tetrahedral ``crystal'' is formed; in the QDC model, the mesh is dodecahedral and a cubic ``crystal'' is formed instead.

Using a mean-field approximation, I have obtained fully analytic results for the thermodynamic properties of the two models. Besides a number of insulating phases, both systems also exhibit a superfluid ground state. In the QDC model, triple coexistence between two solids and the superfluid is superseded by the occurrence of a more stable supersolid phase. Clearly, while the predictions of mean-field theory are generally accurate for an infinite Bose-Hubbard system, deviations will unavoidably be observed in a small system, where no true singularity can occur. However, discrepancies are probably less strong for bosons on a regular spherical mesh, which has equivalent sites and is devoid of natural boundaries. 

To check this expectation for the QCT model, I have diagonalized its Hamiltonian exactly, finding the ground state and a number of ground-state averages. Overall, the exact $T=0$ behavior of the finite system does not depart much from theory, though obviously phase-transition lines are now reduced to simple crossovers; however, clear marks of superfluidity have been detected in the same region of parameters where MF theory predicts it to occur. In conclusion, I hope that the present work may stimulate research aimed at the realization of a new experimental platform for ultracold bosons that is akin to a regular spherical mesh.

\appendix
\section{On the solutions to Eq.~(\ref{eq-30})}

In this Appendix, I show how to determine the eigenvalues of a matrix like (\ref{eq-29}), namely the four real solutions $\lambda$ to the characteristic equation (\ref{eq-30}) with $a+d=b+c\equiv s$.

Upon doing the algebra, the characteristic equation turns out to be:
\ba
(\lambda-a)(\lambda-b)(\lambda-s+a)(\lambda-s+b)-u^2\left[(\lambda-a)(\lambda-b)+(\lambda-s+a)(\lambda-s+b)\right]
\nonumber \\
-v^2\left[(\lambda-a)(\lambda-s+b)+(\lambda-s+a)(\lambda-b)\right]+(u^2-v^2)^2=0\,.
\label{a-1}
\ea
Let us now reshuffle the lhs of (\ref{a-1}) term by term:
\ba
&i)& (\lambda-a)(\lambda-b)(\lambda-s+a)(\lambda-s+b)
\nonumber \\
&&=\frac{1}{16}\left[(2\lambda-s)^2-(2a-s)^2\right]\left[(2\lambda-s)^2-(2b-s)^2\right]
\nonumber \\
&&=\frac{1}{16}\left\{(2\lambda-s)^4-\left[(2a-s)^2+(2b-s)^2\right](2\lambda-s)^2+(2a-s)^2(2b-s)^2\right\}
\nonumber \\
&&=\frac{1}{16}\left\{(2\lambda-s)^4-2\left[(a-b)^2+(a+b-s)^2\right](2\lambda-s)^2+\left[(a-b)^2-(a+b-s)^2\right]^2\right\}\,;
\nonumber \\
\label{a-2}
\ea
\be
ii) -u^2\left[(\lambda-a)(\lambda-b)+(\lambda-s+a)(\lambda-s+b)\right]=-\frac{1}{2}u^2\left[(2\lambda-s)^2+(a+b-s)^2-(a-b)^2\right]\,;
\label{a-3}
\ee
\be
iii) -v^2\left[(\lambda-a)(\lambda-s+b)+(\lambda-s+a)(\lambda-b)\right]=-\frac{1}{2}v^2\left[(2\lambda-s)^2+(a-b)^2-(a+b-s)^2\right]\,.
\label{a-4}
\ee
Plugging Eqs.~(\ref{a-2})-(\ref{a-4}) in (\ref{a-1}), the latter equation becomes:
\ba
&&(2\lambda-s)^4-2\left[(a-b)^2+(a+b-s)^2+4(u^2+v^2)\right](2\lambda-s)^2+\left[(a-b)^2-(a+b-s)^2\right]^2
\nonumber \\
&&+8(u^2-v^2)\left[(a-b)^2-(a+b-s)^2\right]+16(u^2-v^2)^2=0
\label{a-5}
\ea
or equivalently
\ba
&&(2\lambda-s)^4-2\left[(a-b)^2+(a+b-s)^2+4(u^2+v^2)\right](2\lambda-s)^2
\nonumber \\
&&+\left[(a-b)^2-(a+b-s)^2+4(u^2-v^2)\right]^2=0\,.
\label{a-6}
\ea
Upon defining the two quantities $X=(a-b)^2+4u^2$ and $Y=(a+b-s)^2+4v^2$, the lhs of Eq.~(\ref{a-6}) can be factorized as follows:
\ba
&&(2\lambda-s)^4-2(X+Y)(2\lambda-s)^2+(X-Y)^2
\nonumber \\
&=&\left[(2\lambda-s)^2-(X+Y)\right]^2-4XY
\nonumber \\
&=&\left[(2\lambda-s)^2-X-Y+2\sqrt{X}\sqrt{Y}\right]\left[(2\lambda-s)^2-X-Y-2\sqrt{X}\sqrt{Y}\right]
\nonumber \\
&=&\left[(2\lambda-s)^2-\left(\sqrt{X}-\sqrt{Y}\right)^2\right]\left[(2\lambda-s)^2-\left(\sqrt{X}+\sqrt{Y}\right)^2\right]\,.
\label{a-7}
\ea
Hence, two independent second-order equations are obtained from (\ref{a-6}):
\be
(2\lambda-s)^2=\left(\sqrt{X}-\sqrt{Y}\right)^2\,\,\,\,\,\,{\rm and}\,\,\,\,\,\,(2\lambda-s)^2=\left(\sqrt{X}+\sqrt{Y}\right)^2\,,
\label{a-8}
\ee
whose solutions are the searched eigenvalues:
\ba
&&\lambda_1=\frac{1}{2}\left(s-\sqrt{X}+\sqrt{Y}\right)\,;\,\,\,\,\,\,\lambda_2=\frac{1}{2}\left(s+\sqrt{X}-\sqrt{Y}\right)\,;
\nonumber \\
&&\lambda_3=\frac{1}{2}\left(s-\sqrt{X}-\sqrt{Y}\right)\,;\,\,\,\,\,\,\lambda_4=\frac{1}{2}\left(s+\sqrt{X}+\sqrt{Y}\right)\,.
\label{a-9}
\ea
The minimum eigenvalue is solution no.\,3 above.

\section{Self-consistency conditions in the QDC model}
\setcounter{equation}{0}
\renewcommand{\theequation}{B\arabic{equation}}

I here derive the general equations obeyed by MF parameters in the QDC model. These are obtained by first determining the leading eigenvector of the matrix representing the MF grand Hamiltonian on ${\cal B}$. Then, the self-consistency conditions are written in the same way as Eqs.~(\ref{eq-19}).

The MF Hamiltonian is Eq.~(\ref{eq-29}) with real $\phi_A$ and $\phi_B$. Therefore, the eigenvalue equation is as in (\ref{eq-30}), with $a=E_0=48t\phi_A\phi_B+12t\phi_B^2-24V\rho_A\rho_B-6V\rho_B^2,b=E_0+12(2V\rho_A+V\rho_B-\mu),c=E_0+8(3V\rho_B-\mu)$, and $d=E_0+4(6V\rho_A+9V\rho_B-5\mu)$; moreover, $u=-12t(2\phi_A+\phi_B)$ and $v=-24t\phi_B$. The minimum eigenvalue is like in the first line of Eq.~(\ref{eq-31}), but its explicit form in terms of MF parameters is now
\ba
&&E=E_0+2(6V\rho_A+9V\rho_B-5\mu)
\nonumber \\
&&-6\sqrt{(2V\rho_A+V\rho_B-\mu)^2+4t^2(2\phi_A+\phi_B)^2}-4\sqrt{(3V\rho_B-\mu)^2+36t^2\phi_B^2}\,.
\label{b-1}
\ea
Using the latter formula, the linear system for the coordinates $(x,y,z,w)$ of an eigenvector $\left|\psi_E\right>$ of $E$ are:
\be
\left\{
\begin{array}{ll}
-(p_1+p_2)x+q_1y+q_2z=-\left(\sqrt{p_1^2+q_1^2}+\sqrt{p_2^2+q_2^2}\right)\,x & \\
q_1x+(p_1-p_2)y+q_2w=-\left(\sqrt{p_1^2+q_1^2}+\sqrt{p_2^2+q_2^2}\right)\,y & \\
q_2x-(p_1-p_2)z+q_1w=-\left(\sqrt{p_1^2+q_1^2}+\sqrt{p_2^2+q_2^2}\right)\,z & \\
q_2y+q_1z+(p_1+p_2)w=-\left(\sqrt{p_1^2+q_1^2}+\sqrt{p_2^2+q_2^2}\right)\,w &
\end{array}
\right.
\label{b-2}
\ee
with
\ba
&& p_1=3(2V\rho_A+V\rho_B-\mu)\,;\,\,\,\,\,\,q_1=-6t(2\phi_A+\phi_B)\,;
\nonumber \\
&& p_2=2(3V\rho_B-\mu)\,;\,\,\,\,\,\,q_2=-12t\phi_B\,.
\label{b-3}
\ea
Therefore, $\left|\psi_E\right>$ is also eigenvector of a matrix simpler than the original one, with eigenvalue $-\sqrt{p_1^2+q_1^2}-\sqrt{p_2^2+q_2^2}$. When $q_1,q_2\ne 0$, one such vector is:
\be
\left|\psi_E\right>=\left(\frac{p_1+\sqrt{p_1^2+q_1^2}}{q_1}\frac{p_2+\sqrt{p_2^2+q_2^2}}{q_2},-\frac{p_2+\sqrt{p_2^2+q_2^2}}{q_2},-\frac{p_1+\sqrt{p_1^2+q_1^2}}{q_1},1\right)\,.
\label{b-4}
\ee
Now we are ready to impose self-consistency, which eventually leads to:
\be
\rho_A=\frac{1}{2}-\frac{p_2}{2\sqrt{p_2^2+q_2^2}}\,;\,\,\,\,\,\,\rho_B=\frac{1}{2}-\frac{p_1}{2\sqrt{p_1^2+q_1^2}}\,;\,\,\,\,\,\,\phi_A=-\frac{q_2}{2\sqrt{p_2^2+q_2^2}}\,;\,\,\,\,\,\,\phi_B=-\frac{q_1}{2\sqrt{p_1^2+q_1^2}}\,,
\label{b-5}
\ee
or explicitly:
\ba
&& 1-2\rho_A=\frac{3V\rho_B-\mu}{\sqrt{(3V\rho_B-\mu)^2+36t^2\phi_B^2}}\,;\,\,\,\,\,\,
1-2\rho_B=\frac{2V\rho_A+V\rho_B-\mu}{\sqrt{(2V\rho_A+V\rho_B-\mu)^2+4t^2(2\phi_A+\phi_B)^2}}\,;
\nonumber \\
&& \phi_A=\frac{3t\phi_B}{\sqrt{(3V\rho_B-\mu)^2+36t^2\phi_B^2}}\,;\,\,\,\,\,\,\phi_B=\frac{t(2\phi_A+\phi_B)}{\sqrt{(2V\rho_A+V\rho_B-\mu)^2+4t^2(2\phi_A+\phi_B)^2}}\,.
\label{b-6}
\ea
Probably the simpler method to solve these four coupled non-linear equations in four unknowns is to minimize a suitable non-negative function constructed in such a way as to vanish when the (\ref{b-6}) are all fulfilled. This is easy to do numerically with a computer. By this method I have drawn the supersolid boundaries in Fig.\,5.

\end{document}